\documentclass[prb,a4paper,twocolumn,floatfix,showpacs,showkeys,amsmath,amssymb,nobibnotes,altaffilletter]{revtex4}

\usepackage{graphicx}
\usepackage{pslatex}
\usepackage{xspace}
\usepackage{subfigure}

\newcommand{\pcbm}{[6,6]-phenyl-C$_{61}$ butyric acid methyl ester\xspace}

\begin{document}

\preprint{}

\title{Origin of the Efficient Polaron Pair Dissociation in Polymer--Fullerene Blends}

\author{Carsten Deibel}\email{deibel@physik.uni-wuerzburg.de}
\affiliation{Experimental Physics VI, Julius-Maximilians-University of W{\"u}rzburg, 97074 W{\"u}rzburg, Germany}

\author{Thomas Strobel}
\affiliation{Experimental Physics VI, Julius-Maximilians-University of W{\"u}rzburg, 97074 W{\"u}rzburg, Germany}

\author{Vladimir Dyakonov}
\affiliation{Experimental Physics VI, Julius-Maximilians-University of W{\"u}rzburg, 97074 W{\"u}rzburg, Germany}
\affiliation{Bavarian Centre for Applied Energy Research (ZAE Bayern), 97074 W{\"u}rzburg, Germany}

\date{\today}

\begin{abstract}
The separation of photogenerated polaron pairs in organic bulk heterojunction solar cells is the intermediate but crucial step between exciton dissociation and charge transport to the electrodes. In state-of-the-art devices, above $80$\% of all polaron pairs are separated at fields of below $10^7$V/m. In contrast,  considering just the Coulomb binding of the polaron pair, electric fields above $10^8$V/m would be needed to reach similar yields. In order to resolve this discrepancy, we performed kinetic Monte Carlo simulations of polaron pair dissociation in donor--acceptor blends, considering delocalised charge carriers along conjugated polymer chain segments. We show that the resulting fast local charge carrier transport can indeed explain the high experimental quantum yields in polymer solar cells.
\end{abstract}



\maketitle

Polymer--fullerene solar cells have seen steady improvements in terms of performance in the course of the last years. Bulk heterojunction devices showing $5$--$6$\% power conversion efficiency have been published last year,\cite{green2009review} which is only possible by achieving internal photon conversion efficiencies of beyond $80$\%. The light is mainly absorbed in the conjugated polymer, generating singlet excitons, which diffuse towards the interface to the electron accepting fullerene derivative.\cite{hwang2008} The ultrafast electron transfer to the acceptor molecule is very efficient.\cite{sariciftci1992} Electron and hole are now spatially separated but still Coulomb bound. Already a few years ago, the importance of optimizing the morphology of the donor--acceptor phases in these bulk heterojunction devices for an efficient polaron pair separation was shown empirically. A major breakthrough to achieve this was the selection of a suitable solvent.~\cite{shaheen2001} However, even with an optimized phase separation, it was the common understanding that the dissociation was mainly driven by the external electric field. In this paper, the origin of the efficient polaron pair separation in bulk heterojunction solar cells is reconsidered. 

The dissociation of polaron pairs is generally expected only when electric fields exceeding $10^8$V/m are applied.\cite{peumans2004}  In a bulk heterojunction solar cell, however, the internal field in the relevant fourth quadrant of the current--voltage characteristics reaches its maximum value of often less than $10^7$V/m under short circuit conditions. Despite the electric field being more than one order of magnitude smaller,  internal quantum efficiency of $80$\% under short circuit are determined experimentally. This large discrepancy is particularly surprising as the polaron pair lifetime was recently determined to be on the order of only 10ns,\cite{veldman2008} seemingly making the polaron pair dissociation even more unfavorable and challenging. As the classical models such as Onsager~\cite{onsager1938} and Braun~\cite{braun1984} cannot predict these high yields with lifetimes so short, effective lifetimes in the microsecond range have typically been used in modelling and macroscopic simulations.~\cite{deibel2008a}

In order to find the additional driving force assisting the geminate pair separation, Arkhipov et al.\ proposed the analytic dark dipole model, based on the experimental finding of vacuum level shifts at the heterojunction of oligomer/fullerene bilayers.\cite{arkhipov2003} We point out that optical investigations on spin coated polymer films show that the polymer chains were preferentially oriented in parallel to the substrate, with dipole moments laying almost parallel to the chains.\cite{campoy-quiles2005} This effect has been shown to be proportional to the molecular weight,\cite{koynov2006} indicating that a remaining net dipole moment is expected to be present in many conjugated polymers. Another report~\cite{peumans2004} applies a Monte Carlo simulation to show that a large initial separation of the constituents of polaron pairs yields higher dissociation yields, although an additional space charge region had to be assumed to explain the experimental photocurrents of bilayer solar cells. Hot polaron pairs being formed after exciton dissociation, with excess kinetic energy available for the dissociation process, were proposed as model picture. A further approach to improve the dissociation is the high local mobility within ordered domains in the otherwise disordered donor--acceptor blends, as studied for instance by Groves et al.~\cite{groves2008} However, while the high experimental yield is almost reached, this is due to using large donor and acceptor grains, and polaron pair recombination rates orders of magnitude lower than determined experimentally.\cite{veldman2008} 

In this paper, we present kinetic Monte Carlo simulations of hopping transport of Coulomb bound polarons in an energetically disordered donor--acceptor blend system. We investigate the geminate pair dissociation, and how it is influenced by a charge carrier delocalization due to extended polymer chains. For the first time, it is shown that the very efficient polaron pair separation responsible for the high performance of state-of-the-art organic solar cells can be explained by conjugation lengths in the range of 7--10 units.

The simulation space was a simplified bulk heterojunction system, a simple cubic lattice of $100$ units length sandwiched between two electrodes for carrier extraction. The lattice, with a constant spacing of 1nm, had an area of $30\times30$ units with periodic boundary conditions. A model for extended, hole transporting polymer chains---similar to the slithering snake approach in Ref.~\cite{frost2006}---was implemented. The charge carriers within a single segment (the effective conjugation length $CL$ of the polymer chain), given by a fixed number of monomer units, were assumed to be delocalized. For instance, for a polymer chain consisting of 4 monomer units, $CL=4$, delocalization was implemented by placing partial charges (here, quarter charges) on each of these four units. According to the 1nm distance between neighboring sites, this segment is 4nm long, although mostly not straightened. To leave this polymer chain segment, the whole charge can hop `as one' to the next segment, where it spreads again. As the charge transport on a segment (intra-chain transport) is  much faster than the hopping process between chains (inter-chain hopping), the former is assumed to be instantaneous. Charge transport between segments was implemented using the Miller--Abrahams hopping rate,
\begin{equation}
	\nu_{ij} = \nu_0 \exp\left(-2\gamma r_{ij} \right)
		\left\{ \begin{matrix} 
			\exp \left( -\frac{\Delta E_{ij}}{kT} \right) 	& , \Delta E_{ij} > 0 \\ 
			1 								& , \Delta E_{ij} \le 0
		\end{matrix} \right. ,
		\label{eqn:MA}
\end{equation}
where $\nu_0=10^{13}$s$^{-1}$ is the attempt-to-escape frequency and $\gamma=5\cdot 10^9$m$^{-1}$ the inverse localization radius. $r_{ij}$ denotes the hopping distance, and $\Delta E_{ij}$ the energy difference of site i and site j, also accounting for the external electric field and the Coulomb force between all charge carriers and mirror charges. Alternatively to hopping rate~Eqn.~\ref{eqn:MA}, also a polaronic hopping rate could have been used,\cite{parris2001} which might change minor details of the simulation. Half of all sites was defined as electron accepting and transporting molecules, such as fullerene derivatives, each occupying just a single site. Energetic disorder of the density of states was assumed, with a width of the Gaussian distribution of 75meV for the polymer donor and 60meV for the fullerene acceptor, according to our measurements on poly(3-hexyl thiophene) (P3HT) and \pcbm (PCBM).\cite{baumann2009unpublished} We considered ten polaron pairs simultaneously, each consisting of one positive polaron on a donor site and one negative polaron on an acceptor site, yielding a carrier concentration of about $10^{17}$cm$^{-3}$. 

At least 200 runs were performed for each parameter set. We consider a polaron pair separated if at least one constituent is extracted at an electrode, thus escaping geminate and nongeminate recombination. In addition to considering $CL$, we replaced the polaron pair lifetime $\tau_f$ by an effective $\tau_\text{eff} > \tau_f$. Thus, we can account for a reduced recombination probability potentially induced by additional effects such as interface dipoles, which are of theoretical\cite{arkhipov2003} and experimental relevance.\cite{campoy-quiles2005} 

\begin{figure}
	\includegraphics[width=7.5cm]{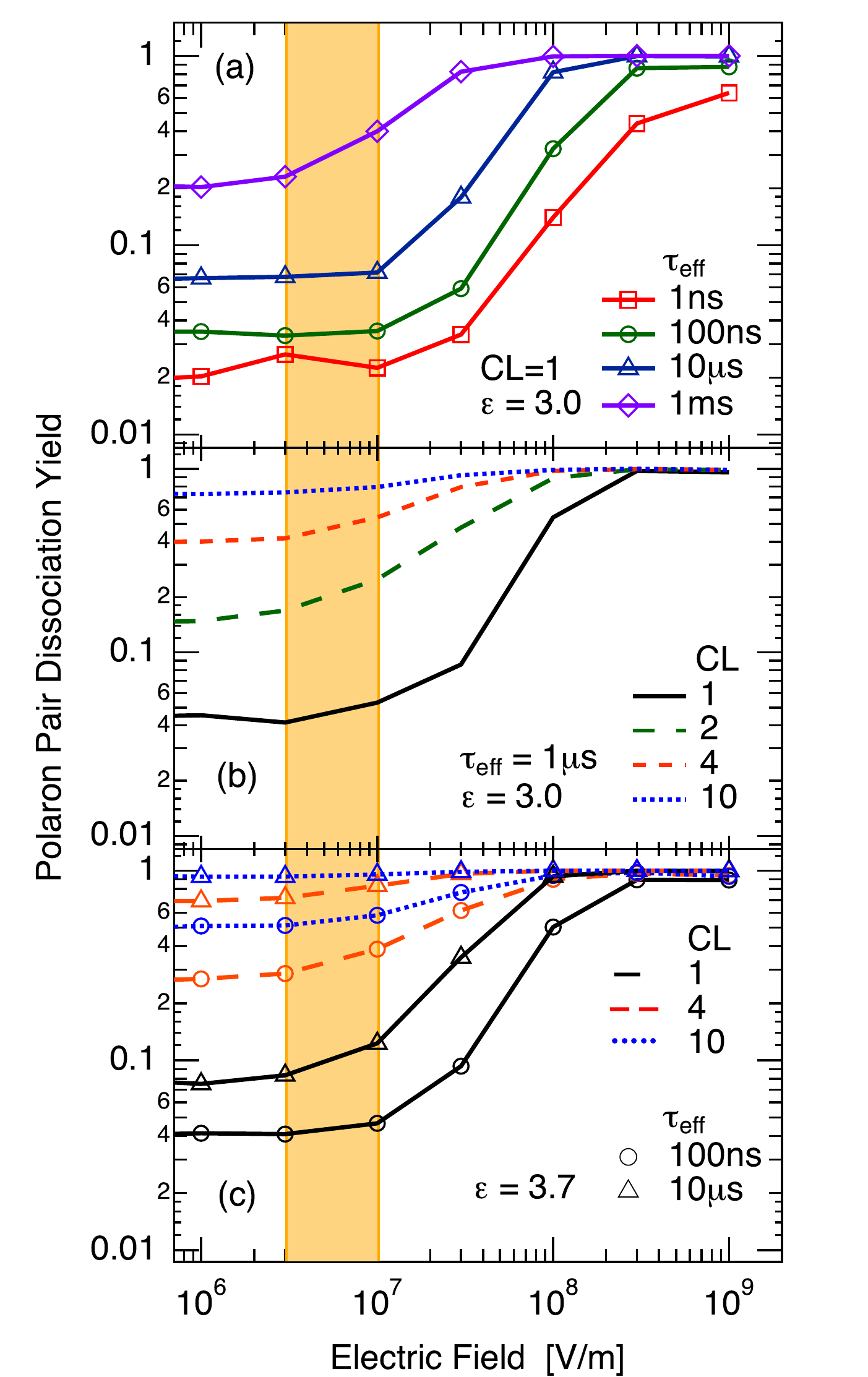}
	\caption{(Color Online) Polaron pair dissociation yield in 1:1 donor--acceptor blends at $300$K. The yellow rectangle denotes the typical internal field under short-circuit conditions of organic solar cells. (a)  Variation of the effective lifetime $\tau_\text{eff}$ at a constant donor $CL$ of 1. The efficient experimental dissociation at low electric fields cannot be reached (dielectric constant $\epsilon=3.0$). (b) Variation of the conjugation length $CL$, with  $\tau_\text{eff}$ fixed at $1\mu$s (dielectric constant $\epsilon=3.0$). (c) Simultaneous variation of $\tau_\text{eff}$ and $CL$, with the dielectric constant chosen as $3.7$. For long polymer chain segments, the high experimentally observed yields are replicated.%
	\label{fig:sim}}
\end{figure}

The influence of the effective lifetime on the polaron pair separation yields at a constant donor conjugation length of one is presented in Fig.~\ref{fig:sim}(a). A lower recombination rate leads to a higher separation yield. Meanwhile, even at effective lifetimes as long as $1$ms, the high experimental dissociation efficiency at low electric fields cannot be explained. The variation of $CL$, and its influence on the geminate pair dissociation, is shown in Fig.~\ref{fig:sim}(b) for a fixed effective lifetime of $1\mu$s. As expected, longer chains lead to an improved separation yield, and show a more positive impact on the yield at low fields as compared to the reduced recombination rate. A longer conjugation length of the donor leads to an increasing local charge carrier mobility and a larger initial electron--hole distance, as the positive charge is delocalised within the conjugated segment. Therefore, the hole can escape the mutual Coulomb attraction more easily, and also needs less hops during its escape. 

The influence of a simultaneous variation of effective lifetime and $CL$ is shown in Fig.~\ref{fig:sim}(c). We point out that the typical conjugation length in polymers used in organic bulk heterojunction solar cells, such as regioregular poly(3-hexyl thiophene), is around $7$--$10$ monomer units.\cite{holdcroft1991} For this regime, the highly efficient polaron pair dissociation process in organic solar cells---as expected from experiment---is for the first time also achieved in Monte Carlo simulations when assuming effective lifetimes between $100$ns and $10\mu$s. The prerequisite for this high efficiency is the physically justified delocalisation of positive polarons within conjugated polymer chain segments.

Veldman et al.\cite{veldman2008} experimentally determined the polaron pair lifetime in a polyfluorene--fullerene blend to about 4ns. Values for the more common poly(3-hexyl thiophene)--fullerene mixture have to our knowledge never been determined. The remaining discrepancy between the effective lifetime of 100ns or more to the experimental lifetime $\tau_f$ can be due to other mechanisms reducing the polaron pair recombination yield. We define $\tau_\text{eff}=\tau_f / p_{rr}$, where $p_{rr}\le 1$ denotes the reduced recombination probability. For $\tau_\text{eff}=1\mu$s and $\tau_f=4ns$, $p_{rr}=1/250$. Considering the Miller--Abrahams hopping rate, Eqn.~\ref{eqn:MA}, this could either correspond to an energetic barrier for recombination of $\Delta E=kT\ln(250)=138$meV at room temperature---e.g., due to dark interface dipoles similar to the considerations of Arkhipov\cite{arkhipov2003}---or to a reduction of the tunneling term for recombination. The latter could be explained by a weakened wavefunction overlap between the donor and acceptor molecule; a reduction of the recombination probablility by factor $250$ is achieved by increasing the inverse localisation radius by factor $\ln (250)=5.5$. Concerning the potential influence of dipoles at the donor--acceptor interface on the geminate pair separation, we found it to be not significant (not shown). This is due to the system being a bulk heterojunction, with surface normals pointing in all directions, thus cancelling each other on longer ranges beyond one nm. The short range influence of the dipoles on sub-nm scale could only be considered by an increased effective lifetime, which shows a clear impact on the dissociation efficiency. Two other potential means of increasing the polaron pair dissociation is the high interchain mobility due to lamellae as reported for poly(3-hexyl thiophene),\cite{sirringhaus1999} or the influence of image force effects if donor and acceptor inhibit different dielectric constants.\cite{szmytkowski2009} Thus, although the impact of the high donor conjugation length on the polaron pair separation efficiency is dominant by far, additional effects driving the dissociation are plausible and expected by us.

In order to relate our simulations to an established model, we calculated the separation yield by the Onsager\cite{onsager1938}--Braun\cite{braun1984} (OB) theory, where the dissociation probability $P(F)$ is given by
\begin{eqnarray}
	P(F) & = & \frac{k_d(F)}{k_d(F) + k_\text{eff}} = \frac{\kappa_d(F)}{\kappa_d(F) + (\mu \tau_\text{eff})^{-1}} .
	\label{eqn:P}
\end{eqnarray}
$F$ is the electric field, $k_\text{eff}=\tau_\text{eff}^{-1}$ the effective polaron pair recombination rate to the ground state, $\mu$ the sum of electron and hole charge carrier mobility. The field dependent dissociation rate $k_d(F)=\mu \kappa_d(F)$ is given by
\begin{equation}
	k_d(F) = \frac{3\gamma}{4\pi r_{pp}^3} \exp\left( - \frac{E_b}{kT} \right) \frac{J_1\left( 2 \sqrt{-2b}  \right)}{\sqrt{-2b}}
	\label{eqn:kd}
\end{equation}
where $\gamma=q\mu/\epsilon\epsilon_0$ is the Langevin recombination factor,\cite{pivrikas2005a} $r_{pp}$ is the initial polaron pair radius, $E_b\propto 1/r_{pp}$ is the Coulombic binding energy of the pair, $kT$ the thermal energy, $J_1$ the Bessel function of order one, and $b=e^3F/(8\pi\epsilon\epsilon_0(kT)^2)$. $e$ is the elementary charge, and $\epsilon\epsilon_0$ the effective dielectric constant of the organic semiconductor blend.  

\begin{figure}
	\includegraphics[width=7.5cm]{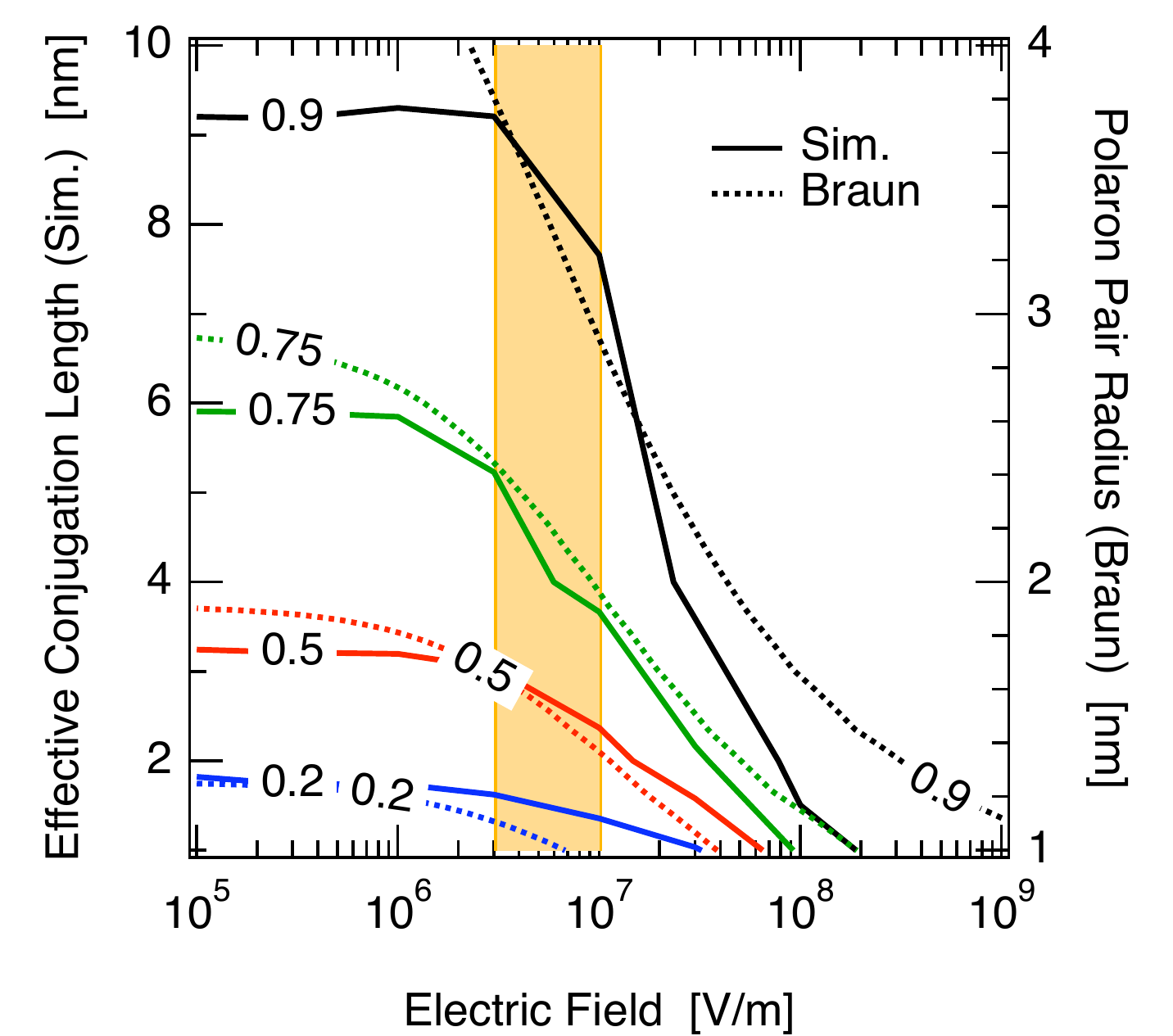}%
	\caption{(Color Online) Polaron pair dissociation yield with Monte Carlo simulation (solid lines) and the OB model (dashed lines) in $1:1$ donor--acceptor blends at $300$K  (dielectric constant $\epsilon=3.0$). The contours are denoted with the yield values. The yellow rectangle denotes again the typical internal field under short-circuit conditions. %
	\label{fig:sim-vs-braun}}
\end{figure}

We considered a spatial distribution of polaron pair separations, as suggested by the Blom group.\cite{mihailetchi2004a} The dissociation yields by Monte Carlo simulation and OB model at $300$K are shown in Fig.~\ref{fig:sim-vs-braun}. For the OB model, we set the $\mu\tau_\text{eff}$-product to $7\cdot 10^{-14}$m$^2$/V---for instance, $10^{-7}$m$^2/$Vs$\cdot700$ns---assuming $CL=2.5r_{pp}$, as the polymer chains are not straightened due to entanglement. 

Comparing the results of the Monte Carlo simulation with the OB model, as shown in Fig.~\ref{fig:sim-vs-braun}, we found a qualitatively good agreement for dissociation yields between 20 and 75\%. The Monte Carlo simulation reaches separation yields of above $90$\% for conjugation lengths beyond 9 units even at low electric fields. Here, the OB theory shows its limitations, as it cannot describe this behaviour. The reason is that the theory does not account for hopping transport and energetic or spatial disorder. 

Recently, experimental findings of the Janssen group in terms of an efficient free polaron generation---performing photoinduced absorption measurements on blends of fluorene copolymers with PCBM---led the authors to propose extended, nanocrystalline fullerene clusters which could yield a high local charge carrier mobility, orders of magnitude beyond the macroscopically determined values.\cite{veldman2008} Using these high local mobilities in the OB model, the authors achieved good fits of their experimental data, assuming cluster sizes of at least 15nm diameter. Solid-state nuclear magnetic resonance relaxometry investigations\cite{mens2007} performed on composites of P3HT and PCBM, however, lead to the conclusion that the fullerene derivative forms amorphous, not nanocrystalline phases.~\cite{mens2009} Illustrating the ongoing debate, Campoy-Quiles et al.\ disagree, pointing out that PCBM can form ordered domains.\cite{campoy-quiles2008} On the other hand, the extended effective conjugation length found in polymers is well-known.\cite{holdcroft1991}  Thus, we propose that the high local on-chain mobility of the conjugated polymer\cite{hoofman1998} due the extended conjugation length is---at least to a significant degree---responsible for the very efficient geminate pair dissociation mechanism in organic solar cells. We point out that the charge transport in the potentially nanocrystalline fullerene phases as well as along semicrystalline lamellae regions found in some polymers such as P3HT\cite{sirringhaus1999} can further increase the effect of fast interchain transport along conjugated polymer chains considered in our simulations.

In conclusion, by performing kinetic Monte Carlo simulations in donor--acceptor blend systems relevant for organic bulk heterojunction solar cells, we found that the highly efficient polaron pair dissociation can be explained by delocalised charge carriers within conjugated segments of the polymer chain. The resulting local charge carrier mobility is much larger than the macroscopic one. Together with the reduced Coulomb attraction due to the accordingly increased intial polaron pair radius, the high on-chain mobility is essential to explain the high polaron pair separation yield. Assuming polymer conjugation lengths between 4 and 10 monomer units, we could simulate 60 to 90\% dissociation yield at moderate fields of below $10^7$V/m, as found in working organic solar cells. These values correspond to recent experimental findings. We qualitatively fit our simulation results with the Onsager--Braun model. Our results show that this model is not well suited to describe polaron pair dissociation at low and high fields, but works well in the intermediate range. We point out that the polaron pair separation is only weakly field dependent in the working regime of organic solar cells, with a finite, significant separation yield at zero internal field. This has important consequences for the modelling and optimisation of organic photovoltaic devices.


C.D. thanks Stijn Verlaak for interesting discussions, and Peter Adriaensens and Sylvain Chambon for making unpublished results available. The current work is supported by the Bundesministerium f{\"u}r Bildung und Forschung in the framework of the GREKOS project (Contract No.~03SF0356B).

\end{document}